\documentstyle [12pt] {article}

\catcode`@=12
\newcommand{\mrm}[1]{\;\mbox{\rm #1}}
\newcommand{\beq}{\begin{equation}}
\newcommand{\eeq}{\end{equation}}

\newcommand{\bea}{\begin{eqnarray}}
\newcommand{\eea}{\end{eqnarray}}

\newcommand{\Eq}[1]{Eq.~(\ref{#1})}

\newcommand{\ea}{{\it et al.}}
\newcommand{\ib}{{\it ibid.\ }}

\newcommand{\plet}[1]{{ Phys. Lett. }{\bf #1}}
\newcommand{\pr}[1]{{ Phys. Rev. }{\bf #1}}
\newcommand{\prlet}[1]{{ Phys. Rev. Lett. }{\bf #1}}
\newcommand{\zp}[1]{{ Z. Phys. }{\bf #1}}

\def\lsim{\mathrel{\vcenter{\hbox{$<$}\nointerlineskip\hbox{$\sim$}}}}

\begin{document}
\thispagestyle{empty}
\begin{flushright} DESY 98-167\\UCRHEP-T239\\November 1998\
\end{flushright}
\vspace{0.5in}
\begin{center}
{\Large \bf Baryogenesis with Scalar Bilinears\\}
\vspace{0.7in}
{\bf Ernest Ma$^1$, Martti Raidal$^{2}$, and Utpal Sarkar$^{2,3}$\\}
\vspace{0.2in}
{$^1$ \sl Department of Physics, University of California\\}
{\sl Riverside, California 92521, USA\\}
\vspace{0.1in}
{$^2$ \sl Theory Group, DESY, D-22603 Hamburg, Germany\\}
\vspace{0.1in}
{$^3$ \sl Physical Research Laboratory, Ahmedabad 380 009, India\\}
\vspace{0.7in}
\end{center}
\begin{abstract}

We show  that if  a  baryon  asymmetry  of  the universe is  generated
through  the out-of-equilibrium   decays   of heavy scalar   bilinears
coupling to  two  fermions  of the  minimal    standard model,  it  is
necessarily  an asymmetry conserving  $(B-L)$ which   cannot survive  
past  the electroweak phase transition because of sphalerons.  We then show 
that a surviving  $(B-L)$ asymmetry may be  generated if  the heavy scalars
decay into two   fermions, \underline  {and  into  two light  scalars}
(which may  be detectable at hadron colliders).   We list all possible
such   trilinear  scalar  interactions,  and    discuss  how our   new
baryogenesis scenario  may   occur naturally in  supersymmetric  grand
unified theories. 

\end{abstract}

\newpage
\baselineskip 24pt

One of the major  successes of grand unified  theories (GUTs) seems to
be the generation of baryon asymmetry of the universe.  After Sakharov
\cite{sakh}  pointed out the three conditions required for
baryogenesis, the first  realization of this  proposal was found  in
GUTs \cite{gutbar}. However, it was later  recognized that the generated
baryon asymmetry conserves $(B-L)$ and is therefore washed away by the
sphaleron-induced,  fast baryon-number violating processes  \cite{krs}
before the electroweak phase transition. 

Restricting ourselves to the fermion content of the standard model (SM), 
we first prove that $(B-L)$ conservation of the 
baryon asymmetry, generated in GUTs through heavy particle decays 
to known fermions only, is a generic feature of any theory.
We then propose a new mechanism for baryogenesis in GUTs
in which a $(B-L)$ asymmetry is generated via
heavy scalar bilinear decays into two fermions and two lighter scalars. 
In this scenario the required  $CP$ violation  comes from the 
interference between the tree-level and one-loop self-energy diagrams. 
We classify all possible trilinear operators of the scalar bilinears
which can contribute to this type of baryogenesis.
We demonstrate that in a wide class of supersymmetric (SUSY) GUTs,
the new  baryogenesis mechanism occurs naturally. 
A generic feature of these scenarios is the existence of light scalars.
For example in some SUSY GUTs, there are pseudo-Goldstone-type bilinears
whose masses are given by seesaw-type relations and 
may be as low as ${\cal O}(1)$ TeV,
giving rise to detectable signatures at future collider experiments.
In particular,
observation of an excess of same-sign lepton pairs or $s$-channel diquark
resonances at the Fermilab Tevatron or the CERN Large Hadron Collider (LHC) 
would strongly support  this proposed
baryogenesis scenario with scalar bilinears.

In spite of the tremendous successes of the SM, there are
now definite experimental indications for physics beyond it. 
With the positive evidence of neutrino masses in atmospheric
 \cite{atm} and solar neutrino \cite{sol} as well as  LSND
\cite{lsnd} experiments, it becomes apparent that we have to extend the SM.
One important approach to understand the new physics beyond the SM is 
to study possible new particles whose existence may be indicated
by the particle content of the SM. 
In the SM the quarks and leptons 
transform under the $SU(3)_C \times SU(2)_L \times U(1)_Y$ gauge group as
$(u_i, d_i)_L \sim (3,2,1/6),$ $u_{iR} \sim (3,1,2/3),$ 
$d_{iR} \sim (3,1,-1/3);$
$(\nu_i, l_i)_L \sim (1,2,-1/2),$ $l_{iR} \sim (1,1,-1),$
where $i=1,2,3$ is the generation index, and there is only one
doublet Higgs scalar, 
$(\phi^+, \phi^0) \sim (1,2,1/2),$
which couples $\overline {(u_i,d_i)}_L$ to $u_{jR}$ and $d_{jR}$, as well as 
$\overline {(\nu_i,l_i)}_L$ to $l_{jR}$.  
However, other scalars which transform as bilinear combinations of the SM
fermions (listed in Table 1) are of great interest. 
There are several scenarios in which new scalar bilinears are added
to explain the masses of  neutrinos.
Dileptons, leptoquarks and diquarks 
inevitably occur in all interesting GUTs \cite{gutrev}.  
They are classified and their phenomenology has been studied in
comprehensive works \cite{frank,ma2}.
In the following we show that they are also important for the generation of
a baryon asymmetry of the universe.

\begin{table}
\begin{center}
\begin{tabular}{|c|c|c|c|c|c|}
\hline
Representation & Notation& $qq$ & 
$\bar q \bar l$ & $q \bar l$ & $ll$  \\
\hline
$(1,1,-1)$ &$\chi^-$ & & & & X  \\
$(1,3,-1)$ &$\xi$     & & & & X  \\
$(1,1,-2)$ &$L^{--}$& & & & X  \\
\hline
$(3^*,1,1/3)$ &$Y_a$& X & X & &  \\
$(3^*,3,1/3)$ & $Y_b$& X & X & & \\
$(3^*,1,4/3)$ & $Y_c$& X & X & &  \\
$(3^*,1,-2/3)$ &$Y_d$& X & & &  \\
\hline
$(3,2,1/6)$ &$X_a$& & & X & \\
$(3,2,7/6)$ &$X_b$& & & X &  \\
\hline
$(6,1,-2/3)$ &$\Delta_a$& X & & &  \\
$(6,1,1/3)$ &$\Delta_b$& X & & &  \\
$(6,1,4/3)$ &$\Delta_c$& X & & &  \\
$(6,3,1/3)$ &$\Delta_{L}$& X & & &   \\
\hline
\end{tabular}
\caption{Scalar bilinears which can 
take part in the generation of baryon asymmetry of the universe.}
\end{center}
\end{table}

To generate a baryon asymmetry it is necessary to have \cite{sakh} 
$(i)$ baryon number violation, $(ii)$ $C$ and $CP$ violation, and $(iii)$ 
out-of-equilibrium conditions.  In early works it was noticed that
baryogenesis is possible in GUTs because there exist new
gauge and Higgs bosons, whose decays violate baryon number. 
The quarks and leptons are put into a single chiral 
representation, implying  mixing of leptoquarks with diquarks. 
As a result, when these heavy particles (say $X$) decay into two quarks
and into a quark and an antilepton, the baryon and lepton numbers are broken
\cite{kolb}.
For $CP$ violation this mechanism requires two heavy gauge or Higgs bosons,
$X$ and $Y$, each of which should have two decay modes,
\begin{eqnarray}
X \to A + B^* , &~~~{\rm and}~~~& X \to C + D^*\,, \nonumber \\
Y \to A + C^* , &~~~{\rm and}~~~& Y \to B + D^*\,, \nonumber 
\end{eqnarray}
so that there exist one-loop vertex corrections to these decays.
The required $CP$ violation occurs due to the interference  
between tree and loop diagrams.
As required by the out-of-equilibrium  condition, 
masses of these particles must satisfy 
\begin{equation}
\Gamma_X < H = 1.7 \sqrt{g_*} {T^2 \over M_P} \hskip .3in {\rm at}~
T = M_X \,,
\label{hubble}
\end{equation}
where, $\Gamma_X$ is the decay rate of the heavy particle $X$; $H$ is
the Hubble constant; $g_*$ is the effective number of massless degrees of 
freedom; and $M_P$ is the Planck scale.
 
In specific GUT scenarios such as $SU(5)$ and $SO(10)$, $(B-L)$ is either 
a global or a local symmetry respectively.  Hence the asymmetry
generated by the above mechanism
is $(B-L)$ conserving \cite{gutrev}. 
When the scalar or vector bosons decay only into fermions, 
any attempt to generate a $(B-L)$ asymmetry leads
to its large suppression in all these models.  Only in models
with a right-handed neutrino, such as $SO(10)$, is it possible to generate 
a $(B-L)$ 
asymmetry after the $(B-L)$ symmetry is broken at some high  scale, 
so that the right-handed neutrinos become massive and since they are Majorana 
fermions, their decays violate lepton number \cite{yanagida}.
Since we are not concerned with any
fermion beyond the SM, this scenario falls
outside the scope of this article. 

The baryon asymmetry generated in the above scenarios 
by the interactions which conserve $(B-L)$ is 
washed out by sphaleron processes \cite{krs} effective at temperatures 
$10^2\mrm{GeV}\lsim T\lsim 10^{12}\mrm{GeV}$.
We shall now prove that 
this is a generic property of the baryon asymmetry generated 
by the above described mechanism, when the decay products are fermions
only, which belong to the SM (not extending it to 
include a right-handed Majorana neutrino). This follows from an 
operator analysis, which was done to show that the minimal scenarios 
of proton decay conserve $(B-L)$ \cite{wein}.
For definitness we consider scalars $X$ and $Y$, but obviously the result 
generalizes also to vectors.

Let us start from the Lagrangian giving the decays of $X$ and $Y$,
\begin{equation}
{\cal L} = f_{xab} \bar{A} B X + f_{xcd} \bar{C} D X + 
f_{yac} \bar{A} C Y + f_{ybd} \bar{B} D Y\,. 
\end{equation}
As per our assumption, there are no new fermions in addition to those
present in the SM and the scalars $X,$ $Y$ decay into only quarks and leptons.
To obtain a nonzero $CP$ violation from the interference
between tree and vertex  diagrams, 
we require $X$ and $Y$ to be distinct from each
other and to have different decay modes. This implies $B$ and $C$ to
be distinct. In the SM one can then write down all 
possible combinations of $A$, $B$, $C$, and $D$, with $X$ and $Y$, and 
find out the decay modes of $X$ and $Y$. On the other hand, since
the out-of-equilibrium condition and the nonvanishing 
of the absorptive part of the loop integral require these scalars
$X$ and $Y$ to be much heavier than the fermions, we can integrate 
them out inside the loop and write down the diagrams in terms 
of the four-fermion effective operators of the SM, as shown in Fig.~1.

\begin{figure}
\vskip 3.5in\relax\noindent\hskip -.1in\relax{\includegraphics{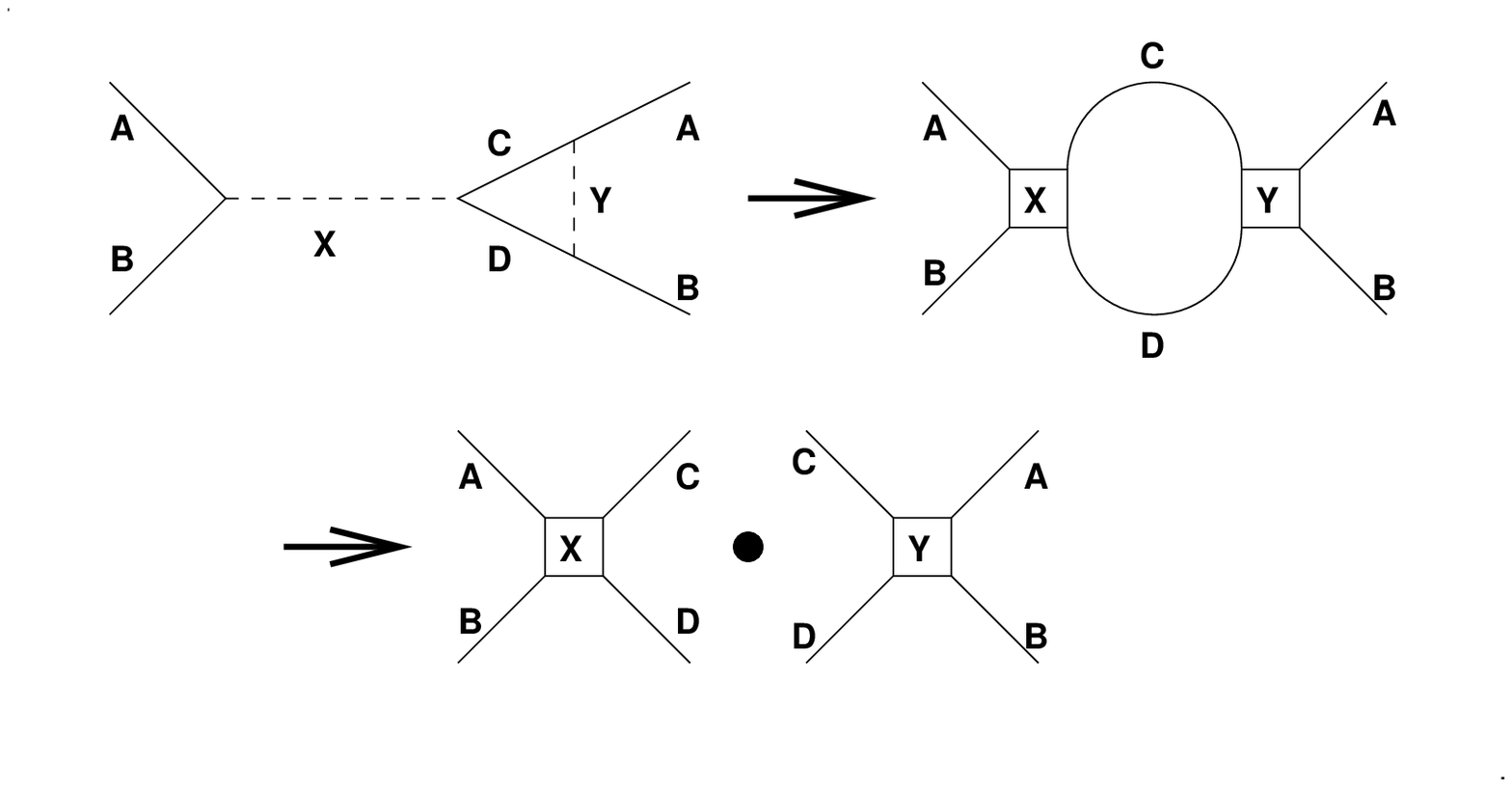}}
\caption{Interference of effective four fermion operators which generates
baryon asymmetry.}
\end{figure}

This simple but crucial step allows us to use existing knowledge on SM 
four-fermion operators for baryon number violation which have been studied 
extensively in the literature \cite{wein}.  It  was
found that all these operators conserve $(B-L)$ to the lowest 
order. Any $(B-L)$ violating operator will be suppressed by 
${<\phi>^2 / M_{GUT}^2}$ compared to the $(B+L)$ violating 
operators. In models with an intermediate symmetry breaking scale
or with new Higgs scalars at some intermediate scales, this 
suppression factor may be softened a little, but still strong 
enough to rule out any possibility of generating enough baryon
asymmetry of the universe. 
On the other hand, any four-fermion operator which 
violates only lepton number requires all the fermions to be the same; 
hence it cannot generate the required $CP$ asymmetry. 
Therefore a $(B-L)$ asymmetry, needed to survive 
the sphaleron processes, is impossible with the SM four-fermion operators. 

In the considered scenario, one can in principle also have 
the self-energy-type diagrams with the fermions in the loop for 
generating the $CP$ asymmetry.  However, in this case,
after integrating out the heavy scalars, the effective diagrams 
in terms of the four-fermion operators are exactly the same as in the 
vertex-correction case, so the conclusion is unchanged.
As long as the heavy scalars decay only into fermions, the generated 
asymmetry always conserves $(B-L)$. This generic feature is a 
consequence of the SM fermionic content.

We now show how a $(B-L)$ asymmetry can be generated in GUTs if there are 
both heavy and light scalar bilinears.  This is a generalization of a 
recently proposed scenario of leptogenesis 
\cite{prl}, where each of two heavy scalars decays into two fermions 
and into two light scalars.  Low-energy effective operators now contain
two fermions and two scalar bilinears. The  required $CP$ 
violation for baryogenesis comes entirely 
from an interference between  the tree-level decay and the self-energy
corrections \cite{prl}, and there are no one-loop vertex corrections, 
as would be the case with Eq.~(2). 

Consider the scalars $S_{1,2}$, each of which can decay into 
two fermions $\psi_1 + \psi_2$ and into two scalars $Z_1 + Z_2$.
If the $(B-L)$ quantum numbers for the two decay modes are different,
these processes  violate $(B-L)$. 
The Lagrangian describing  these interactions is of the form
\begin{equation}
{\cal L} = M_a^2 S_a^\dagger  S_a + \left( f_a \overline{\psi_1^c} 
\psi_2 S_a^\dagger + \mu_a Z_1 Z_2 S_a^\dagger + h.c. \right) \,, 
\end{equation}
where the fermions $\psi_{1,2}$ and the scalars $Z_{1,2}$ are assumed
to be much lighter than $S_{1,2}.$  This is then exactly analogous to 
Eq.~(14) of Ref.~[13] and we can simply use the formalism developed there 
to obtain the $(B-L)$ asymmetries generated by the tree level decays of 
the physical states approximating $S_{1,2}$ and their interference with
the one-loop self-energy diagrams \cite{prl}, which is given by
\begin{equation}
\delta_a \simeq \Delta (B-L)
{{Im \left[ \mu_1 \mu_2^* f_1^* f_2 \right]} \over 
{16 \pi^2 (M_1^2 - M_2^2)}} \left[ {{ M_a} \over 
\Gamma_a}  \right],
\label{delta}
\end{equation}
where the width $\Gamma_a$ is given by $( |\mu_a|^2 + M_a^2 |f_a|^2 ) 
/ (8 \pi M_a)$.

Let $M_1 > M_2$, then as the universe cooled down to below $M_1$, most of 
$S_1$ would decay away.  However, the asymmetry so created would be erased 
by the $(B-L)$ nonconserving interactions of $S_2$.  Hence only the 
subsequent decay of $S_2$ at $T < M_2$ would generate a $(B-L)$ asymmetry 
which would pass through the electroweak phase transition unscathed. 
If $S_2$ is heavy enough to satisfy the out-of-equilibrium condition
$\Gamma_a < H$ of Eq.~(1), then the
final baryon asymmetry is approximately given by~\cite{kolb} 
$\delta_B \sim \delta_2 / (3 g_*)$.  The desired value 
of $\delta_B \sim 10^{-10}$ may thus be obtained with a variety of scalar 
masses and couplings.

At energies below the heavy scalar $S_{1,2}$
masses, lower bounds of which can be obtained from Eq.~(1),  
any $(B-L)$ violating effective operator of the form 
\bea
{\cal O}_{(B-L)} \equiv [\psi_1 \psi_2 Z_1^\dagger Z_2^\dagger]
\label{operator}
\eea 
can generate the baryon asymmetry. 
In the SM there is only one Higgs doublet
scalar $\phi$ which is supposed to be light. Hence
there can be only one $(B-L)$ violating effective operator of the required 
form, {\it i.e.} $ l_i l_j \phi \phi $, which can be obtained from 
the SM particles. 
This operator has been studied in the literature extensively. It 
contains all the scenarios of neutrino masses and leptogenesis \cite{ma1}.
For example, it can be induced by
the triplet bilinear $\xi$ in Table 1 generating a
lepton asymmetry of the universe \cite{prl}.
This operator may also originate from heavy Majorana neutrinos \cite{yanagida}.

In GUTs where the scalar bilinears listed in Table 1 occur, there
are many other possibilities to form  dimension-five 
operators of the type given by \Eq{operator} which violate lepton and baryon
numbers. As all the scalar bilinears couple to ordinary fermions,
the classification of the two-scalar-two-fermion 
baryon-asymmetry generating operators in GUTs
reduces to that of all possible $(B-L)$ violating trilinear operators of the
scalar bilinears, as shown in Table 2. 
{}From this list, we see that the first two trilinear scalar operators, 
${\cal O}_{1,2},$ give rise to the well-known dimension-five operator 
$ l_i l_j \phi \phi$ \cite{wein}.  The rest occur in GUTs such as $SO(10)$ and 
$E_6$, as will be demonstrated below.  Note the interesting fact that 
$|\Delta (B-L)| = 2$ in all cases.

\begin{table}
\begin{center}
\begin{tabular}{|c|c|c|c|c|c|}
\hline 
Operators&$B-L$&Operators&$B-L$&Operators&$B-L$ \\
\hline 
\phantom{${\cal O}_1 = \mu_1 \phi \phi \chi^- $ }&&
\phantom{${\cal O}_1 = \mu_1 \phi \phi \chi^- $ }&&
\phantom{${\cal O}_1 = \mu_1 \phi \phi \chi^- $ }& \\
${\cal O}_1 = \mu_1 \phi \phi \chi^-$ & -2 &$ {\cal O}_2 =\mu_2 \phi 
\phi \xi$ &  -2 &${\cal O}_3 = \mu_3 \chi^- \chi^- L^{++}$ & -2 \\
${\cal O}_4 = \mu_4 \xi \xi L^{++} $ & -2 &$ {\cal O}_5 = \mu_5 Y_a 
Y_c^\dagger \chi^+ $ & 2 &$ {\cal O}_6 = \mu_6 Y_d Y_a Y_a 
$ & 2 \\$
{\cal O}_7 = \mu_7 Y_d Y_b Y_b $ & 2 &$ 
{\cal O}_8 = \mu_8 Y_c Y_d Y_d $ & 2 &$
{\cal O}_9 = \mu_9 Y_b Y_c^\dagger {\xi}^\dagger $ & 2 \\$
{\cal O}_{10} = \mu_{10} Y_a Y_d^\dagger {\chi^-} $ & -2 &$
{\cal O}_{11} = \mu_{11} Y_b Y_d^\dagger {\xi} $ & -2 &$
{\cal O}_{12} = \mu_{12} Y_c Y_d^\dagger L^{--}  $ & -2 \\$
{\cal O}_{13} = \mu_{13} X_b X_a^\dagger \chi^- $ & -2 &$
{\cal O}_{14} = \mu_{14} X_b X_a^\dagger \xi $ & -2 &$
{\cal O}_{15} = \mu_{15} X_a X_b Y_c^\dagger $ & 2 \\$
{\cal O}_{16} = \mu_{16} X_a \phi Y_d $ & 2 &$
{\cal O}_{17} = \mu_{17} X_a \phi^\dagger Y_a $ & 2 &$
{\cal O}_{18} = \mu_{18} X_a \phi^\dagger Y_b $ & 2 \\$
{\cal O}_{19} = \mu_{19} X_a X_a Y_a^\dagger $ & 2 &$
{\cal O}_{20} = \mu_{20} X_a X_a Y_b^\dagger $ & 2 &$
{\cal O}_{21} = \mu_{21} X_b Y_d \phi^\dagger $ & 2 \\$
{\cal O}_{22} = \mu_{22} \Delta_a Y_a Y_a $ & 2 &$
{\cal O}_{23} = \mu_{23} \Delta_a Y_b Y_b $ & 2 &$
{\cal O}_{24} = \mu_{24} \Delta_a \Delta_b \Delta_b $ & 2 \\$
{\cal O}_{25} = \mu_{25} \Delta_c \Delta_a \Delta_a $ & 2 &$
{\cal O}_{26} = \mu_{26} \Delta_c Y_d Y_d $ & 2 &$
{\cal O}_{27} = \mu_{27} \Delta_b X_a^\dagger X_a^\dagger $ & -2 \\$
{\cal O}_{28} = \mu_{28} \Delta_L Y_b Y_d $ & 2 &$
{\cal O}_{29} = \mu_{29} \Delta_b Y_a Y_d $ & 2 &$
{\cal O}_{30} = \mu_{30} \Delta_a Y_d Y_c $ & 2 \\$
{\cal O}_{31} = \mu_{31} \Delta_c X_a^\dagger X_b^\dagger $ & -2 &$
{\cal O}_{32} = \mu_{32} \Delta_L X_a^\dagger X_a^\dagger $ & -2 &$
{\cal O}_{33} = \mu_{33} \Delta_L \Delta_L \Delta_a $ & 2 \\$
{\cal O}_{34} = \mu_{34} \Delta_a^\dagger \Delta_b \chi^- $ & -2 &$
{\cal O}_{35} = \mu_{35} \Delta_a^\dagger \Delta_L \xi $ & -2 &$
{\cal O}_{36} = \mu_{36} \Delta_b^\dagger \Delta_c \chi^- $ & -2 \\$
{\cal O}_{37} = \mu_{37} \Delta_L^\dagger \Delta_c \xi $ & -2 &$
{\cal O}_{38} = \mu_{38} \Delta_a^\dagger \Delta_c L^{--} $ & -2 &
& \\
\phantom{${\cal O}_1 = \mu_1 \phi \phi \chi^- $ }&&
\phantom{${\cal O}_1 = \mu_1 \phi \phi \chi^- $ }&&
\phantom{${\cal O}_1 = \mu_1 \phi \phi \chi^- $ }& \\
\hline 
\end{tabular}
\end{center}
\caption{Trilinear scalar operators which can contribute to the
baryon asymmetry of the universe.}
\end{table}

To exemplify the general discussion
we shall now consider a large class of SUSY $SO(10)$ GUTs.
The $SO(10)$ symmetry may be broken down to the SM symmetry through
several intermediate steps which include the Pati-Salam
$SU(4)_C \times SU(2)_L \times SU(2)_R$ and/or
$SU(2)_L \times SU(2)_R\times U(1)_{B-L}$ symmetries \cite{lr}.
It has been shown \cite{goran,chacko} that at these intermediate stages, 
the requirement of stabilizing the charge-conserving vacuum 
after breaking the supersymmetry
introduces higher-dimensional operators to the theory.
The resulting low-energy theory is the  R--parity 
conserving minimal supersymmetric SM plus light diquark, leptoquark, and
dilepton states,
which obtain masses via seesaw-type relations.

In the supersymmetric limit and in the absence of the
nonrenormalizable terms, the superpotential 
of a minimal SUSY Pati--Salam intermediate theory \cite{ps}
has a complexified $U(30)$
symmetry that operates on $SU(2)$ triplet, $SU(4)$ tenplet superfields. 
After the neutral components of the triplets 
acquire  vacuum expectation values at the scale $M_R$, thus breaking the 
symmetry, a $U(29)$ complexified symmetry remains,  giving
rise to 118 massless fields, 18 of which get masses from the $D$ terms. 
Inclusion of the higher-dimensional effective terms necessary to conserve
the electric charge leads thus to a total of 50 complex pseudo-Goldstone bosons
with masses $m_{pG}\sim M_R^2/M_{P},$ where $M_P$ is the Planck scale. 
For $M_R$ as high as ${\cal O}(10^{10})$ GeV, the pseudo-Goldstone-type 
diquarks, leptoquarks, and dileptons may have masses 
${\cal O}(1)$ TeV. 
More details can be found in  Ref. \cite{chacko}.

Let us consider one of the choices which leaves one $Y_b$ field as light 
as ${\cal O}(1)$ TeV. Then, for example, 
the operator ${\cal O}_{23}$ in Table 2  implies that some of
the heavy $\Delta_a$ could generate a baryon asymmetry of the universe.
Even though the left-right symmetry breaking scale is around $10^{10}$ GeV, 
the $\Delta_a$ can be much heavier than this mass scale. The out-of-equilibrium
condition implies that these fields are as heavy as $10^{13}$ GeV.
Their decay modes into $Y_b + Y_b$ and into $d^c + d^c$ 
violate baryon number as well as $(B-L)$.  Hence a baryon asymmetry
of the universe can be generated according to the mechanism discussed before. 
Since this is a $(B-L)$ asymmetry, it will not be washed away by the sphaleron 
processes. Note that the light $Y_b$ alone can be assigned definite 
global quantum numbers, and hence they do not wash away the generated 
baryon asymmetry;
their Yukawa interactions should satisfy constraints derived in 
Ref. \cite{frank,ma2}.

An important feature of our new baryogenesis mechanism in general, and 
the discussed SUSY GUT scenario in particular,  is that
the light scalar bilinear fields  can lead to  detectable signatures 
at the Fermilab Tevatron or the CERN LHC. The most interesting among these
are the $s$-channel resonance processes mediated by diquarks \cite{ma2}.
They may result in resonance production of light dijets or distinct 
final states such as $tc$ or $tt.$ The leptonic decays of two top quarks
provide same-sign dilepton final states which have very little SM 
background.  At the Tevatron, the $s$-channel production
is sea-quark suppressed and diquark masses up to only ${\cal O}(1)$ TeV 
are testable in the $tc,$ $tt$ channels, but at the LHC, diquark masses
as high as ${\cal O}(10)$ TeV can be probed \cite{ma2}.
Therefore, any possible signal of this type detected  at hadron colliders 
will lend support to the proposed baryogenesis mechanism.

To summarize, we have shown  that a $(B-L)$ asymmetry cannot be generated 
in GUTs if the new heavy gauge bosons or scalar bilinears decay only into
the SM fermions. 
As a result the baryon asymmetry of the universe generated 
by this type of mechanism cannot survive to the present day 
because it would have been washed away by the sphaleron processes. 
We then show that it is possible to generate a $(B-L)$ asymmetry in GUTs 
using scalar bilinear decays into known fermions and into 
light scalars.  We have classified 
all possible operators of the scalar bilinears
which can contribute to this baryogenesis mechanism.
As an example we have demonstrated that 
the proposed baryogenesis mechanism occurs naturally
in a wide class of SUSY GUTs based on the $SO(10)$ gauge symmetry. 
The light scalar bilinears may lead to clear  
detectable experimental signatures at colliders, especially
in the discussed SUSY GUTs where they are pesudo-Goldstone bosons with 
mass of ${\cal O}(1)$ TeV. 

~\vskip 0.2in
\begin{center} {ACKNOWLEDGEMENT}
\end{center}

We thank W. Buchm\"uller for useful comments on the manuscript and 
F. Vissani for discussions. 
MR and US acknowledge financial support from the Alexander von 
Humboldt Foundation and hospitality of DESY Theory Group.
EM also acknowledges the hospitality of DESY (where this work was initiated) 
and of CERN (where this work was completed). 
The work of EM was supported in part by the U.~S. Department of Energy 
under Grant No.~DE-FG03-94ER40837.

\newpage
\bibliographystyle{unsrt}

\end{document}